\documentstyle[12pt,epsf]{article}
\newcommand{\vb}[1] {\mbox {\bf{#1}}}

\begin{document}

\title{DETUNING INDUCED EFFECTS:
SYMMETRY-BREAKING BIFURCATIONS
IN DYNAMIC MODEL OF ONE-MODE LASER}
\author{Alexei D. Kiselev\thanks{E-mail: adk@kid.ti.chernigov.ua}\\[3mm]
Department of Pure and Applied Mathematics,\\
Chernigov Technological Institute,\\
Shevchenko St., 95, 250027 Chernigov, UKRAINE\\[4mm]}
\date{}
\maketitle

\begin{abstract}

The concept of broken
symmetry is used to study bifurcations of equilibria and dynamical
instabilities in dynamic model of one-mode laser (nonresonant complex
Lorenz model) on the basis of modified Hopf theory.
It is shown that an invariant set of stationary points bifurcates into an
invariant torus (doubly-periodic branching solution). Influence of the
symmetry breaking on stability of branching solutions is investigated as a
function of detuning. The invariant torus is found to be stable under the
detuning exceeds its critical value, so that dynamically broken symmetry
results in the appearance of low frequency Goldstone-type mode. If the
detuning then goes downward and pumping is kept above the threshold,
numerical analysis reveals that after a cascade of period-doublings the
strange Lorenz attractor is formed at small values of detuning. It is found
that there are three different types of the system behaviour as pumping
increases depending on the detuning. Quantum counterpart of the complex
Lorenz model is discussed.

\vspace*{2cm} PACS numbers: 42.60Mi, 42.65Sf, 05.45+b, 82.20Mj

\end{abstract}

\newpage

\section{INTRODUCTION}

Nonlinear dynamics of laser systems, especially, those that exhibit so-called
chaotic behavior  has been the subject of major interest during the last two
decades (see, for example, \cite{AM1,GB}). In the semiclassical approximation,
when quantum correlation effects are disregarded, description of dynamical
instabilities in lasers is shown to be closely related to the theory of
nonequilibrium phase transitions in dissipative dynamical systems \cite{VL}.
From this standpoint, of particular interest are the systems that can be
reduced to the well-known models of the dynamical system theory. One of the
models is the famous Lorenz equations:
$\dot{X}=\sigma(-X+Y),\,\dot{Y}=rX-Y+XZ,\, \dot{Z}=-bZ-XY$
that were originally derived in \cite{LR} and have been studied intensively
from the end of the seventies (see, for example, [5-9] and references therein).
The above system (in what follows it will be referred as the
real Lorenz model for $X(t)$, $Y(t)$ and $Z(t)$ are real-valued functions)
was obtained from a set of hydrodynamic equations in the three-mode
approximation to describe the convective motion of a layer of fluid that is
warmer at the bottom than at the top, so that $\sigma$ is the Prandtl number;
$r$ (controlling parameter) is the Rayleigh number and is proportional to the
temperature difference; $b$ depends on geometrical properties of the fluid
layer.

In order to clarify the relevance of Lorenz-type models to  laser physics let
us consider the simplest one-mode laser equations taken in the following form
\cite{HK1}:

\begin{equation}
\left\{
\begin{array}{crl}
\dot{b}& = &-(\kappa+i\omega)\cdot b-ig\cdot\alpha \\
\dot{\alpha}& = &-(\gamma+i\omega_{a})\cdot\alpha+ig\cdot b\cdot S\\
\dot{S}& = &(d_{0}-S)/T-4g\cdot \Im( \alpha\cdot \overline{b})
\end{array} \right.
\end{equation}

where $b$ is the dimensionless complex amplitude of the electromagnetic field
mode; $\omega$ ($\kappa$) is the frequency (the relaxation constant) of the
mode; $\alpha$ is the dimensionless transition dipole matrix element; $S$
is the inversion of the atomic level populations; $\omega_{a}$ is the
frequency of the atomic transition; $d_{0}$ is the parameter characterizing
the intensity of pumping; $g$ is the coupling constant; $\gamma$ ($T^{-1}$) is
the transverse (longitudinal) relaxation  constant.

Eq.(1) is derived within the framework of semiclassical approach to dymanics
of the system that constitutes a number of two-level atoms (atomic subsystem)
interacting with the one mode of electromagnetic field (field subsystem), so
that $b$ is the averaged field anihilation operator descriptive of the
coherent part of radiation. Note that the equations can be obtained from
Heisenberg operator equations for Dicke-type hamiltonian taken in the
rotating wave approximation after neglecting of quantum fluctuations and
assuming no dependence on spatial variables \cite{HK1,AE}. In addition, the
result should be supplemented with the relaxation terms as well as the term
proportional to the intensity of pumping (see Sec. 5 for more details on
quantum models behind Eq.(1)).

After making the substitutions: $$t\rightarrow\gamma t,\,b=\gamma X/(2g),
\,\alpha=i S_{0} Y/2,\,S=d_{0}+S_{0} Z,\,S_{0}=\gamma
\kappa/ g^{2}$$
and going over to the representation of interaction the system (1)
can be rewritten as a complex Lorenz model \cite{CL}:

\begin{equation}
\left\{
\begin{array}{crl}
\dot{X}& = &\sigma (-(1+i\Delta)\cdot X+Y) \\
\dot{Y}& = &-(1-i\Delta)\cdot Y+r \cdot X+X\cdot Z\\
\dot{Z}& = &-b\cdot Z-\Re( X\cdot \overline{Y})
\end{array} \right.
\end{equation}

where $\sigma=\kappa/\gamma,\,r=d_{0}/S_{0},\, b=(\gamma T)^{-1},\,
\Delta=(\omega_{a}-\omega)/(\kappa+\gamma)$ is the frequency detuning. Note that
$X(t)$ and $Y(t)$ are complex-valued functions, so that the system (2)
consists of five real equations.

In the case of exact resonance when $\Delta = 0$, though the complex Lorenz
model is appeared to differ from the real one in some respects, it was shown
that basically there is no difference between dynamics  of the system (2) and
one of the real Lorenz model \cite{CL}.

In this paper our main purpose is to study detuning induced effects
in dynamics of the complex Lorenz model on the basis of bifurcation analysis.
The latter means that we deal with stability and existence of certain
branching solutions depending on the intensity of pumping. In other words,
$r$ is assumed to be a control parameter. By contrast to
the real Lorenz model,the system (2) has a continuous symmetry group
(Lie group of rotations in complex planes), so that we approach the problem
within the unified concept of symmetry breaking. It should be emphasized that
the reason behind all qualitatively new effects discussed in the paper is
precisely the symmetry and $\Delta$ can be served as a quantity to measure
influence of dynamically broken symmetry on bifurcating solutions and their
stability.

In order to discuss the effects in more specific terms, let us outline some
relevant results for the real Lorenz model (most of them hold at $\Delta = 0$).

As a preliminary we comment on stability and bifurcations of
equilibria (steady states). The null steady state corresponding to the
spontaneous emission regime (no coherent radiation) is given by $X= Y= Z= 0$
and is asymptotically stable at $r< r_{0}= 1$. This solution loses its
stability at $r= r_{0}$ and there are two asymptotically stable bifurcating
steady states $X=Y= \pm\sqrt{b(r-1)}, \, Z= r-1$ provided that $r>r_{0}$.
At this stage we have the stationary points bifurcated from the origin at
$r=r_{0}$ indicating that the regime of spontaneous emission is changed to the
initiation of laser generation (convective flow for the problem of \cite{LR}).
This bifurcation produces qualitative changes in phase portrait of the system.
The analogy between transformation of such kind, that can be regarded as a
nonequilibrium phase transition, and a second order phase transition in
thermodynamic systems led to the synergetic concept of a phase transition
\cite{HK2}.

If $\sigma > b+1$, there is another critical value of the control
parameter $r$:
$\displaystyle r_{c}=\frac{\sigma\cdot(\sigma+b+3)}{\sigma-b-1}$,
such that the above two stationary points become exponentially unstable at
$r>r_{c}$. Note that the linearized operator governing stability of the
solutions in question (Liapunov's first theorem) has a pair of complex
conjugate eigenvalues with negative real parts in the neighborhood of $r_{c}$.
These eigenvalues are pure imaginary at $r= r_{c}$. According to the Hopf
bifurcation theory \cite{HP}, it follows that a new branching time
periodic solution can be expected to appear (Hopf bifurcation). Stability of
this solution is determined by the Floquet exponents: the solution is stable
(unstable) if it appears supercritically (subcritically). In the case under
consideration the bifurcation was found to be subcritical \cite{MM}. It
implies that the time periodic solution is unstable when the pumping exceeded
its critical value given by $r_{c}$.

One of the most striking features of the real Lorenz model is the appearance
of so-called strange Lorenz attractor instantaneously on passing the
control parameter $r$ through its critical value $r_{c}$ ("drastic"
route to chaos). The term 'strange attractor'(or 'chaotic attractor') is
commonly used for an attracting set that has a rather complicated structure
and each trajectory within the attractor is exponentially unstable. There is
a number of different quantities to measure the complexity (stochasticity) of
the attractor structure:  capacity (fractal dimension), information dimension,
Hausdorff and Liapunov (Kaplan-Yorke formula) dimensions, K-entropy and so on.

Detailed description of how the strange Lorenz attractor forms is beyond the
scope of this paper. In brief, this can be understood as being due to the
occurrence of a homoclinic orbit in the system: as $r$ passes
through the value at which the homoclinic 'explosion' takes place, a strange
invariant set of trajectories is produced, including an infinite number of
periodic orbit \cite{SH,SP,NL}. Note that, in addition, the real Lorenz
model is known to exhibit period-doubling \cite{RB}, intermittency \cite{MP}
and hysteresis \cite{FM} in various ranges of its parameter space.

The paper is organised as follows:

In Sec. 2 it is shown that due to the symmetry the null equilibrium state
of the complex Lorenz model (spontaneous emission) bifurcate into an
invariant set of stationary points at $r=r_{0}=1+\Delta^{2}$. Stability of
the equilibrium states is studied as a function of detuning. It is found that
under $\sigma > b+1$ there is a critical value of the control parameter
$r$ (pumping intensity), $r_{c}$, such that the states of the invariant set
become exponentially unstable at $r>r_{c}$ and $r_{c}$ is an increasing
function of $\Delta^{2}$.

In Sec. 3 analytical power series Hopf technique is extended on the system
invariant under the action of a continuous symmetry group to construct
bifurcating solutions and to investigate their stability in the vicinity of
$r=r_{c}$ at $\Delta \ne 0$. Due to the symmetry breaking the bifurating
solution is appeared to be doubly-periodic at $\Delta \ne 0$. It means that
nonzero detuning results in the appearance of low-frequency Goldstone-type
mode related to the motion along an orbit of the group. Moreover, the broken
symmetry is found to affect stability of the branching solution.

In Sec. 4 the results of numerical analysis are discussed. Dependence of the
relevant Floquet exponent on $\Delta$ is calculated. It is obtained that
there is a critical detuning, $\Delta_{c}$, such that the bifurcating doubly-
periodic solution is stable at $|\Delta |> \Delta_{c}$. It leads to the
formation of stable invariant torus. Different routes to chaos depending
on the detuning are duscussed.

Concluding remarks and discussion are given in Sec. 5.
Bosonic three-oscillator quantum system that gives the complex Lorenz model
in the semiclassical approximation (quantum counterpart) is formulated.
\newpage

\section{SYMMETRY OF THE PROBLEM:
STABILITY AND BIFURCATIONS OF EQUILIBRIA}

Taking $r$ as a bifurcation (control) parameter, Eq.(2) can be rewritten
as an autonoumous dynamical system with quadratic nonlinearity:

\begin{equation}
\dot{\vb{x}}=\vb{f}(\vb{x})=L\,\vb{x}+\vb{f}_{2}(\vb{x},\vb{x})
\end{equation}

where
$$\vb{x}=(x_{1},\,x_{2},\,x_{3},\,x_{4},\,x_{5}),\,X=x_{1}+ix_{2},\,
Y=x_{3}+ix_{4},\,Z=x_{5},$$
$$L=D\vb{f}(0)\,
(L_{ij}=\frac{\partial f_{i}}{\partial x_{j}}(0));\;
f_{2}^{i}(\vb{x},\vb{y})=\sum_{n,m}
\frac{\partial^2 f_{i}}{\partial x_{n}\partial x_{m}}\cdot x_{n}\cdot y_{m}$$

Clearly, the system (2) is invariant under the transformations:
$$X\rightarrow\exp(i\phi)\cdot X,\;Y\rightarrow\exp(i\phi)\cdot Y,\;
Z\rightarrow Z,$$
so that Eq.(3) has a continuous symmetry group $G$ of rotations in the
$x_{1}-x_{2}$ and $x_{3}-x_{4}$ planes, $G\sim SU(1)$:

\begin{equation}
\vb{f}(T(\phi)\cdot \vb{x})=T(\phi)\cdot\vb{f}(\vb{x}),\;\;T(\phi)\in G
\end{equation}

Note that there is an additional discrete symmetry in the complex Lorenz
model:
$$Y\rightarrow\overline{X},\;Y\rightarrow\overline{Y},\;
Z\rightarrow Z,\;\Delta\rightarrow -\Delta\, .$$
The latter implies no dependence on the sign of detuning. For brevity, in what
follows $\Delta$ is assumed to be nonnegative.

Eq.(4) gives
\begin{equation}
\vb{f}(T(\phi)\cdot \vb{x}_{st})=0\Rightarrow |D\vb{f}(\vb{x}_{st})|=0,
\end{equation}
where $\vb{x}_{st}$ is a noninvariant equilibrium solution to Eq.(3) and its
orbit, $T(\phi)\cdot \vb{x}_{st}$, produces an invariant set of equilibria.
The second equation in (5) is obtained by differentiating the first one with
respect to $\phi$. Along this line we come to the conclusion that matrix of
the first approximation, that govern linearized stability of $\vb{x}_{st}$, is
degenerate and its null vector  is
$\displaystyle \vb{e}_{0}=K\cdot\vb{x}_{st},$
where $\displaystyle K=\frac{d}{d\phi} T(\phi) |_{\phi=0}$ is the generator
of $G$.

It can be readily shown that the null stationary point related to the regime
of spontaneous emission loses its stability after the pumping intensity
exceeded its critical value
\begin{equation}
r>r_{0}=1+\Delta^2
\end{equation}
and the solution bifurcated from the origin can be taken in the form
\begin{equation}
x_{1}^{st}=x_{3}^{st}=q=\sqrt{b\cdot (r-1-\Delta^{2})},\;
x_{2}^{st}=0,\;x_{4}^{st}=\Delta\cdot q,\;
x_{5}=1+\Delta^{2}-r.
\end{equation}
According to Eq.(5), the stationary point (7) provides an invariant set of
the steady states corresponding to the appearance of coherent light emission
(laser generation). In the first approximation stability of the state (7) is
determined by eigenvalues of the following matrix:

\begin{equation}
D\vb{f}(\vb{x}_{st})=L_{c}=\left(
\begin{array}{ccccc}
-\sigma & \Delta\cdot\sigma & \sigma & 0 & 0 \\
-\Delta\cdot\sigma & -\sigma & 0 & \sigma & 0 \\
1+\Delta^{2} & 0 & -1 & -\Delta & q \\
0 & 1+\Delta^{2} & \Delta & -1 & 0 \\
q & -\Delta\cdot q & -q & 0 & -b \\
\end{array} \right)
\end{equation}

As it can be seen from Eq.(5), $|L_{c}|=0$ and the kernel of $L_{c}$ is
defined by the null vector $\vb{e}_{0}=(0,\,1,\,-\Delta,\,1,\,0)$.
Strictly speaking, we cannot conclude on stability of stationary
point in the first approximation for one of the eigenvalues is zero. So the
underlying arguments need to be modified. Obviously, since there is another
steady state in any neighborhood of a given steady state on the orbit, it
cannot enjoy the property of being asymptotically stable. Suppose that all
other four eigenvalues have negative real parts and sketch a way how
stability of the invariant set can be studied on the basis of symmetry
arguments. First, the set is asymptotically stable if

\begin{equation}
\|\vb{x}(t)-T(\psi(t))\vb{x}_{st} \| \rightarrow 0 \: ,\: t\rightarrow\infty
\end{equation}

where $\vb{x}(t)$ is a solution of Eq.(3) with initial conditions taken in
the vicinity of a stationary point; $\psi(t)$ is determined from the condition
$$\|\vb{x}(t)-T(\psi(t))\vb{x}_{st} \| \rightarrow \min $$
at given $t$, so that the left hand side of Eq.(9) defines the distance from
$\vb{x}(t)$ to the invariant set. The latter can be written in the form:

\begin{equation}
\vb{z}(t)=T(-\psi(t))\vb{x}(t) \perp \vb{e}_{0}
\end{equation}

From Eqs(3, 10) it is not difficult to obtain the equation for $\vb{z}(t)$
and to derive the expression for $\dot{\psi(t)}$. Then the motion along the
orbit can be eliminated from the consideration and  subsequent analysis of the
modified system can be made in the first approximation. Following this line
we arrive at the conclusion that the invariant set is asymptotically stable
in the case under investigation. Some rather straightforward algebra on the
subject was made in \cite{BK} for $\Delta=0$.

If $\sigma>b+1$ (this condition is found to be independent of $\Delta$),
there is a critical value of the bifurcation parameter, such that the
solution in question is unstable at $r>r_{c}$ and $L_{c}$ has a pair of
complex conjugate imaginary eigenvalues
$\lambda_{1}=\overline{\lambda}_{2}=i\alpha$ at $r=r_{c}$.
Equations for $r_{c}$ and $\alpha$ can be derived by
making use of Routh-Hurwitz criteria:

\begin{eqnarray}
p_{1}p_{2}p_{3} & = & p_{1}^{2} +p_{0}p_{3}^{2} \\
\alpha^{2} & = & \frac{p_{1}}{p_{3}},
\end{eqnarray}

where $p_{i}$ are coefficients of the characteristic
polynomial:

\begin{eqnarray}
p_{0} & = & 2\sigma b ( 1+\sigma )(r-1-\Delta^{2}) \\
p_{1} & = & b ( (1+\Delta^{2})\sigma^{2}+3\sigma (r-2)+r) \\
p_{2} & = & (\sigma+1)^{2} +1+r+2 \sigma +\Delta^{2}((\sigma-1)^{2}-b) \\
p_{3} & = & 2\sigma + b + 2.
\end{eqnarray}

Dependence of the ratio $r_{c}$ and $r_{0}$ on detuning is presented in
Fig. 1 at $b =1$ for $\sigma = 3$ (solid line) and $\sigma = 5$ (dashed line).
It is seen that $r_{c}$ is an increasing function of $\Delta$.

In addition, it can be obtained
that $\displaystyle \Re{(\frac{d}{dr}\lambda_{1}|_{r=r_{c}})}>0$ and then, if
not symmetry induced degeneracy of $L_{c}$,  we could meet all conditions for
the occurence of the Hopf bifurcation.

\section{HOPF-TYPE BIFURCATION}

There are different techniques to investigate Hopf bifurcations and stability
of bifurcating time periodic solutions \cite{HP}. In this section we  adopt
the perturbative method, closely related to the analytical technique by Hopf,
to explicitly construct the bifurcating solution in the form of power series
over small parameter $\epsilon$ characterizing amplitude of the solution
in the neighborhood of the bifurcation point $r=r_{c}$ \cite{HP,IJ1}.
Coefficients of the power series can be derived by making use of Fredholm
alternative and linearized stability of the solution, determined by the
Floquet exponents, can be studied on the basis of the factorization theorem
\cite{JS}.

In trying to make analysis along the above line we need to modify the method
to bypass mathematical difficulties arising from the degeneracy of $L_{c}$.
To avoid equations that have no solutions it is assumed that the symmetry is
'spontaneously' broken and the bifurcating solution is taken in the form:

\begin{equation}
\vb{x}=T(\epsilon\theta)\cdot \vb{u}(t)=
T(\epsilon\theta)\cdot (\vb{x}_{st}+\epsilon\cdot \vb{z}(t))
\end{equation}

Substitution of Eq.(17) in Eq.(3) gives the equation for $\vb{u}(t)$ :

\begin{equation}
\dot{\vb{u}}+\epsilon\cdot K \vb{u}=\vb{f}(\vb{u})
\end{equation}

Let us introduce the renormalized frequency
$\Omega(\epsilon)=\alpha/(1+\tau(\epsilon))$,
so that $\vb{z}(s)=\vb{z}(\Omega t)$ is a $2\pi$-periodic
vector-valued function. The equation for $\vb{z}(s)$ reads

\begin{equation}
\alpha\cdot(\dot{\vb{z}}+\dot{\theta}\cdot(\vb{e}_{0}+\epsilon\cdot K\cdot
\vb{z}))
=(1+\tau)\cdot(L_{c}\,\vb{z}+\epsilon\, \vb{f}_{2}(\vb{z},\vb{z}))
\end{equation}

where $L_{c}$ is defined by Eq.(8) and the dot stands for the derivative with
respect to $s$.
The $2\pi$-periodic solution of Eq.(19) can be found in the form of power
series in $\epsilon$:

\begin{equation}
\vb{z}(s)=\sum_{n=0}^{\infty}\vb{z}_{n}(s)\,\epsilon^{n},\:
\tau=\sum_{n=1}^{\infty}\tau_{n} \epsilon^{n},\:
\dot{\theta}=\sum_{n=0}^{\infty}\omega_{n} \epsilon^{n}
\end{equation}

As it is seen from Eqs.(7, 8), the quantity $q=\sqrt{b\cdot (r-1-\Delta^{2})}$
can be conveniently chosen as a bifurcation parameter, so that
\begin{eqnarray}
q & = & q_{c}+\Delta q=q_{c}+\sum_{n=1}^{\infty} q_{n} \epsilon^{n}\\
L_{c} & = & L_{c}^{(0)}+L_{c}^{(1)}\cdot\Delta q\\
\vb{e}_{0} & = & K \vb{x}_{st}=
\vb{e}_{0}^{(0)}+\vb{e}_{0}^{(1)}\cdot\Delta q
\end{eqnarray}
where $L_{c}^{(0)},\,\vb{e}_{0}^{(0)}$ are $L_{c},\,\vb{e}_{0}$ at
$q=q_{c}$. Recall that $L_{c}^{(0)}$ has a pair of pure imaginary
complex conjugate eigenvalues:
$L_{c}^{(0)}\vb{e}_{1}=i\alpha\, \vb{e}_{1},\,
L_{c}^{(0)}\vb{e}_{2}=-i\alpha\, \vb{e}_{2}$. (For brevity, the superscript
(0) will be omitted from the eigenvector notations.) Note that eigenvectors
$\{\vb{e}_{0}^{\ast},...\,\vb{e}_{5}^{\ast}\}$ of the operator adjoint to
$L_{c}^{(0)}$  and $\{\vb{e}_{0},...\,\vb{e}_{5}\}$
are mutually orthogonal:
$\langle\vb{e}_{i}^{\ast},\vb{e}_{j}\rangle=\delta_{ij}$.

According to the standard perturbative technique, insertion of Eqs.(20-23)
into Eq.(19) yields equations to be solved for $\vb{z}_{n}(s)$ successively.
In the case of $n > 0$, the equations can be written in the form:
\begin{equation}
\{\alpha\cdot\frac{d}{ds}-L_{c}^{(0)}\}\,\vb{z}_{n}(s)=
\vb{f}^{(n)}(s)=\vb{f}_{0}^{(n)}+\sum_{k=1}^{2n-2}
[\,\vb{f}_{k}^{(n)}\cdot\exp{iks}+c.c.\,]
\end{equation}

Parameters $q_{n}$, $\omega_{n}$ and $\tau_{n}$ can be found by making use of
Fredholm alternative.
The latter states that Eq.(24) has $2\pi$-periodic solutions if and
only if
\begin{equation}
\langle\vb{e}_{0}^{\ast},\vb{f}_{0}^{(n)}\rangle =0
\end{equation}
\begin{equation}
\langle\vb{e}_{1}^{\ast},\vb{f}_{1}^{(n)}\rangle =0
\end{equation}
Note that Eq.(25) is the solvability condition for time independent part of
$\vb{z}_{n}(s)$ and Eq.(26) is to get rid of so-called secular terms.

In order to reduce remaining arbitrariness in choice of the solutions,
$\vb{z}_{n}(s)$ are subjected to the following additional constraints:

\begin{equation}
\langle\vb{e}_{0}^{\ast},\vb{z}_{n}(s)\rangle=0
\end{equation}
\begin{equation}
\int_{0}^{2\pi}
\langle\vb{e}_{1}^{\ast},\vb{z}_{n}(s)\rangle\exp{(-is)}ds=0,\,n>0
\end{equation}

In the zero-order approximation we have
\begin{equation}
\{\alpha\cdot\frac{d}{ds}-L_{c}^{(0)}\}\,\vb{z}_{0}(s)=
-\alpha\,\omega_{0}\,\vb{e}_{0}
\end{equation}
so that
\begin{equation}
\vb{z}_{0}(s)=A\cdot\vb{e}_{1}\,\exp{(is)}+c.c.,\:\:\omega_{0}=0
\end{equation}
where $A$ is a complex integration constant that is determined by the
initial condition for $\vb{z}_{0}(s)$ and can be eliminated from the
consideration by renormalizing the eigenvector $\vb{e}_{1}$.

After some straightforward calculations the following results can be
obtained:
\begin{equation}
q_{2n+1}=\tau_{2n+1}=\omega_{2n}=0
\end{equation}
\begin{equation}
\alpha\cdot\omega_{1}=2\cdot |A|^{2}\cdot
\langle\vb{e}_{0}^{\ast},\vb{f}_{2}(\vb{e}_{1},\vb{e}_{2})\rangle
\end{equation}
\begin{equation}
\vb{z}_{1}(s)=\vb{b}_{0}+
\{A^{2}\cdot\vb{b}_{1}\cdot\exp{(2is)}+c.c.\}
\end{equation}
\begin{equation}
-q_{2}\cdot\Re(\frac{\partial}{\partial q}\lambda_{c})=\Re(k)
\end{equation}
where $\lambda_{c}$ is the eigenvalue of $L_{c}$, such that
$\lambda_{c}=i\alpha$ at $q=q_{c}$,
\begin{equation}
k=2\cdot |A|^{2}\cdot\{\,
2\cdot\langle\vb{e}_{1}^{\ast},\vb{f}_{2}(\vb{e}_{1},\vb{b}_{0})\rangle+
\langle\vb{e}_{1}^{\ast},\vb{f}_{2}(\vb{e}_{2},\vb{b}_{1})\rangle\,\}-
\alpha\cdot\omega_{1}\cdot\langle\vb{e}_{1}^{\ast},K\vb{e}_{1}\rangle
\end{equation}
and the vectors $\vb{b}_{0},\:\vb{b}_{1}$ are solutions of the equations:
\begin{equation}
\{2i\alpha-L_{c}^{(0)}\}\,\vb{b}_{1}=
\vb{f}_{2}(\vb{e}_{1},\vb{e}_{1})
\end{equation}
\begin{equation}
-L_{c}^{(0)}\,\vb{b}_{0}=
\vb{f}_{2}(\vb{e}_{1},\vb{e}_{2})-
\langle\vb{e}_{0}^{\ast},\vb{f}_{2}(\vb{e}_{1},\vb{e}_{2})\rangle
\cdot\vb{e}_{0}.
\end{equation}

At this stage we get the modification of the Hopf theory with allowance
for the symmetry breaking, so that bifurcating solution is appeared to be
doubly periodic. In other words, there are two types of frequencies:
the basic frequency $\Omega$ and the Goldstone-type low frequency mode
with $\epsilon\dot{\theta}\sim\Delta^2$. (The latter can be inferred from
Eq.(32).) It is of interest to note that another distinctive feature of the
above results is the last term of Eq.(35). This term, being geometrical in
nature, is accounted for the broken symmetry and will be shown to be of
importance to stability analysis.

In the Hopf theory conclusion on stability of the bifurcating solution can be
drawn from Eqs.(34, 35) based on the factorization theorem \cite{HP,IJ1,JS}
that states about the stability depending on the sign of $\Re(k)$: if
$\Re(k)<0$, then bifurcation is supercritical and the time periodic branching
solution is stable at $q>q_{c}$; if $\Re(k)>0$ the solution appears
subcritically.

Note that the factorization theorem should be extended to system invariant
under the action of a Lie symmetry group. For the system under consideration
the theorem can be recovered by making use ansatz that looks like (17) and
real part of the relevant Floquet exponent is proportional to $\Re(k)$. More
details on the subject is given in \cite{AD1}.

As a result, Eqs.(34, 35) are key equations for making conclusion on
stability of the bifurcating invariant torus. In particular, it
implies that the torus is stable at $r>r_{c}$ under $\Re{(k)}<0$.

\section{NUMERICAL RESULTS}

In the previous section we have studied how symmetry of the system affects
Hopf-type bifurcation at $r=r_{c}$. Our findings are:
\begin{enumerate}
\item An invariant set of equilibria bifurcates into an invariant torus. In
other words, the branching solution is time doubly-periodic, so that
Goldstone-type low frequency mode is found to appear due to the symmetry
breaking.

\item It is found that the sign of $\Re{(k)}$ with $k$ defined by Eq.(35)
determines stability of the torus. The last symmetry induced term in Eq.(35)
implies that the broken symmetry affects stability of the branching solution.
\end{enumerate}
Note that the frequency of the Goldstone-type mode as well as the last term of
Eq.(35) tend to zero as $\Delta\rightarrow 0$.

In Fig. 2 are shown dependencies of $\Re{(k)}$ on detuning ($\Delta$)
for $\sigma=5$ (solid line) and $\sigma=10$ (dashed line) at $b=1$. It is
seen that in both cases there is a critical detuning, $\Delta_{c}$, at which
$\Re{(k)}$ changes its sign, so that bifurcation being subcritical at
$\Delta<\Delta_{c}$ becomes supercritical at $\Delta>\Delta_{c}$. The latter
means that an invariant set of equilibria Eq.(7), corresponding to the laser
generation, bifurcates into the stable torus as $r$ passes through $r_{c}$
under $\Delta>\Delta_{c}$.

To get some idea of qualitative changes of  attractor structure in
relation to detuning, there are three trajectories in 3D
$\Re{(X)}-\Re{(Y)}-Z$ space in Figs. 3-5 presented at $\sigma =5,\,b=1$
and $r=r_{c}+0.2$ for various values of $\Delta$. Fig. 3 is clearly revealed
the attractor as an invariant torus at $\Delta=0.5>\Delta_{c}\approx 0.41$,
whereas we have the well-known Lorenz attractor under $\Delta=0$ (Fig. 5).
As is shown in Fig. 4, the intermediate case of $\Delta=0.1$ corresponds to
an entangled structure which is hard to interpret.

One of the ways to clarify the point is to look at the relevant Fourier
spectra. To this end, the Fourier spectra $|X(\omega)|$ and
$|Z(\omega)|$ are calculated at $\Delta=0.5$ (Fig. 6) and
$\Delta=0.1$ (Fig. 7). Notice that $|X(\omega)|^{2}$ is proportional to
the power spectrum of the electromagnetic field.

The Fourier spectrum $|X(\omega)|$ for $\Delta=0.5$, depicted in Fig. 3,
indicates the high frequency peak at $\omega\approx\alpha=5.5$ and the two
intensive low frequency peaks at $\omega=0$ and $\omega\approx 0.04$. Since
the frequency $\omega\approx 0.04$ does not contribute to the spectrum
$|Z(\omega)|$, this peak can be attributed to the Goldstone-type mode.
So, the numerical results are in agreement with ones obtained from the
theoretical analysis of Sec. 3. As far as numerical analysis is concerned, it
should be emphasized that, working with relatively small number of points
(less than 20000), we are not to present the results of high precision
calculations, but our calculations has been made with reasonable accuracy for
investigation of the theoretical predictions qualitatively.

Coming back to Fig. 4 and looking at the Fourier spectra in Fig. 7,
let us recall that, according to the theory of Sec. 3 and Fig. 2, the relevant
Floquet exponent is pure imaginary at $\Delta=\Delta_{c}$. So changing
$\Delta$ from above $\Delta_{c}$, where the invariant 2D torus is stable,
downward ($r$ is fixed) we encounter another bifurcation point at
$\Delta=\Delta_{c}$, and the torus is expected to bifurcate into a 3D
torus, embedded in the 5D phase space of the complex Lorenz model. Computer
simulation confirms this conclusion. Comprehensive analysis of this secondary
bifurcation is beyond the scope of this paper
(some results on the subject were obtained in \cite{IJ2}). Further
decrease of $\Delta$ would result in other bifurcations. Taking into
account that the spectra of Fig. 7 are typical of period doublings, it can be
suggested that  the chaotic attractor forms at relatively small $\Delta$
after a cascade of doublings. The irregular Fourier spectrum at
$\Delta=0.05$, displayed in Fig.  8, clearly indicates chaotic
dynamics of the system.

\section{DISCUSSION AND CONCLUDING REMARKS}

In this paper we have studied some detuning induced effects in dynamical
model of the single-mode laser. The key point of stability analysis and
bifurcation theory, presented in Sec. 2-3, is that dynamical symmetry
breaking must be taken into consideration. It is shown that the symmetry
breaking results in formation of an invariant set of equilibria, which is an
orbit of the stationary point given by Eq.(7), at $r=r_{0}$ and leads to the
appearance of low frequency Goldstone-type mode related to the motion
along the orbit at $r=r_{c}$. Moreover, it is found that there is the
symmetry induced term in Eq. (35), so that stability of the branching
doubly-periodic solution (invariant torus) is affected by
the symmetry breaking.

Coming back to dymanics of the laser system (Eq.(1)),
let us discuss what are the effects that could be
observed experimentally. It should be noticed that, typically, it is
difficult to meet the condition $r>r_{c}$ in a single-mode laser and it
was just a few experiments with gas lasers, where the threshold of
dynamical instability was exceeded \cite{KW,AM}. Based on the results
of the above theory (Sec. 2-3) and numerical analysis (Sec. 4), in
experimental setup of \cite{AM} with homogeneously broadened one-mode
$CO_{2}$ laser one could expect three different types of the system behaviour
as pumping increases:

\begin{itemize}
\item At sufficiently small detuning, $\Delta<\Delta_{0}$,
($\Delta_{0}\approx 0.07$ at $\sigma=5$ and $b=1$) the chaotic attractor
forms abruptly during the passage of $r$ through $r_{c}$;
\item The system undergoes a cascade of doublings before its transition to
chaos at $\Delta_{0}<\Delta<\Delta_{c}$;
\item If $\Delta>\Delta_{c}$, the system does not reveal chaotic behaviour
even if $r\approx 10-20\, r_{c}$.
\end{itemize}

This gives an insight into why just an oscillatory instability was observed
for single-mode operation of the laser in \cite{KW,AM}. The effect
can be attributed to the off-center pumping that was used to ensure the
single-mode operation, so that the detuning $\Delta$ was greater than its
critical value $\Delta_{c}$.

From the other hand, recently the real Lorenz equations has been employed to
study three parameter kinetics of a phase transition \cite{OK}. The model
was found to represent the main features of a second order phase transition
in the case of real order parameter. It seems to be straightforward to extend
the arguments given in \cite{OK} to the case of complex order parameter, so
that the complex Lorenz model could play an important part in investigation
of the kinetics of a nonequilibrium second order phase transition. Notice
that, according to synergetic approach \cite{HK2}, a phase transition
is realized as a result of mutual coordination between the complex order
parameter ($X$), the conjugate field ($Y$) and the control parameter ($Z$).
So our results can be regarded as an extension of the analogy between
nonequilibrium phase transitions and phase transitions in thermodynamic
systems.

In conclusion, we give some details on a quantum counterpart of the complex
Lorenz model. As it was mentioned in Sec. 1, in the semiclassical
approximation, the well-known N-center Dicke hamiltonian \cite{PY}:

\begin{equation}
H_{D}=\omega\cdot b^{+}b+\frac{\omega_{a}}{2}\cdot\sigma^{z}
+g\cdot (b^{+}\,\sigma^{-}+b\,\sigma^{+})
\end{equation}

$$ \sigma^{\pm}=\sum_{r=1}^{N} \sigma^{\pm}_{r},\;
\sigma^{z}=\sum_{r=1}^{N} \sigma^{z}_{r}$$

can provide the Lorenz equations. Recall that equations of motion must be
supplemented with the relaxation terms as well as the term descriptive of
pumping. As far as the problem of quantum chaos is concerned, Eq.(38)
cannot be considered as an explicit quantum counterpart of Eq.(2). Clearly,
the reason is that relaxation and pumping do not enter Eq.(38).

One way to get rid
of the above shortcoming is to use two oscillator representation for the Pauli
operators that enter Eq.(38) \cite{PY,AK}:
$\sigma^{+}\rightarrow a^{+}_{2}a_{1},\: \sigma^{-}\rightarrow a^{+}_{1}a_{2}$,
where $a^{+}_{i}$  ($a_{i}$)
is the bosonic creation (anihilation) operator of the $i$-th oscillator.
The resulting hamiltonian reads:

\begin{equation}
H=\omega\cdot b^{+}b+\omega_{1}\cdot a^{+}_{1}a_{1}+
\omega_{2}\cdot a^{+}_{2}a_{2}
+g\cdot (b^{+}\, a^{+}_{1}\, a_{2}+b\, a^{+}_{2}\, a_{1})
\end{equation}

The next step is to write master equation for density matrix $\rho$
in the form \cite{WL}:

\begin{equation}
-\dot{\rho}=i[H,\rho]+\gamma_{0}\cdot L_{b}\,\rho +
\gamma_{1}\cdot L_{1}\,\rho+\gamma_{2}\cdot L_{2}\,\rho
\end{equation}

where

$$
L_{b}\,\rho=\{[b^{+},b\rho]-[b,\rho\,b^{+}]\}+\exp{(-\beta\omega)}\cdot
\{[b,b^{+}\rho]-[b^{+},\rho\,b]\},
$$
$$
L_{i}\,\rho=\{[a^{+}_{i},a_{i}\rho]-[a_{i},\rho\,a^{+}_{i}]\}+
\exp{(-\beta_{i}\omega_{i})}\cdot
\{[a_{i},a^{+}_{i}\rho]-[a^{+}_{i},\rho\,a_{i}]\},
$$
$\beta=1/(k_{B}T_{0}),\,\beta_{i}=1/(k_{B}T_{i})$; $k_{B}$ is the Boltzman
constant. It is supposed that the $i$-th oscillator interacts with its
thermostat characterized by a temperature of $T_{i}$ and the thermostats
are statistically independent(more details on systems of such kind can be
found in \cite{GM, AD}). Note that the exact solution of the three oscillator
model (Eq. (39)) was recently derived by making use of the algebraic Bethe
ansatz \cite{BN}.

Assuming that $\langle b a^{+}_{i}a_{j}\rangle\approx\langle
b\rangle\cdot \langle a^{+}_{i}a_{j}\rangle$ and $\Gamma_{1}=\Gamma_{2}$
($\Gamma_{i}= \gamma_{i}\cdot (1-\exp{(-\beta_{i}\omega_{i})})$), Eq.(1) can
be readily derived from Eq. (40). So we have:

\begin{equation} \{b,\,\alpha ,
\, S\} \leftrightarrow \{\langle b\rangle ,\, \langle a^{+}_{2}a_{1}\rangle ,
\, \langle a^{+}_{2}a_{2}-a^{+}_{1}a_{1}\rangle\},
\end{equation}

\begin{equation}
\omega_{a}=\omega_{2}-\omega_{1},\:
\kappa = \gamma_{0}\cdot (1-\exp{(-\beta\omega)}),\:
\gamma = 1/T = 2\,\Gamma_{1},
\end{equation}

\begin{equation}
d_{0}/2 = \langle n \rangle _{2}-\langle n \rangle _{1},
\end{equation}

where $\langle n \rangle _{i}=(\exp{(\beta_{i}\omega_{i})}-1)^{-1}$. Note that
for $d_{0}$ to be positive it is necessary to meet the condition:
$T_{2}/T_{1}>\omega_{2}/\omega_{1}$.

Thus, phenomenological parameters are expressed in terms of microscopic
quantities by Eqs.(41-43) and Eqs.(39-40) yield the explicit quantum
counterpart of the complex Lorenz model that can be employed to study the
problem of quantum chaos \cite{KN}. This work is under progress.

\begin{center}
{\bf ACKNOWLEGMENTS}
\end{center}

Author is grateful to Prof. A.I. Olemskoi for stimulating remarks and
valuable discussion during his stay at Sumy State University.

\newpage

\begin{center}
{\bf FIGURE CAPTIONS}
\end{center}

\begin{description}

\item[Fig. 1] The ratio of the dynamical instability threshold $r_{c}$ to
the laser generation threshold $r_{0}=1+\Delta^{2}$ as a function of detuning
$\Delta$ at $b=1$ for $\sigma = 3$ (solid line) and $\sigma = 5$ (dashed
line). It is seen that $r_{c}$ is an increasing function of $\Delta$.

\item[Fig. 2] Dependence of $\Re{(k)}$ (see Eq.(35)) on $\Delta$ at
$b=1$ for $\sigma=5$ (solid line) and $\sigma=10$ (dashed line). In both of
the cases the coefficient is shown to change its sign at critical value of
the detuning.

\item[Fig. 3] Trajectory in 3D $\Re{(X)}-\Re{(Y)}-Z$
subspace  at $\sigma=5,\,b=1$ and $r=r_{c}+0.2$ for
$\Delta=0.5,\,r_{c}\approx 23.41$. The trajectory is shown to wind up the
stable invariant torus.

\item[Fig. 4] Trajectory in 3D $\Re{(X)}-\Re{(Y)}-Z$
subspace  at $\sigma=5,\,b=1$ and $r=r_{c}+0.2$ for
$\Delta=0.1,\,r_{c}\approx 15.28$.

\item[Fig. 5] The strange Lorenz attractor in 3D $\Re{(X)}-\Re{(Y)}-Z$
subspace  at $\sigma=5,\,b=1$ and $r=r_{c}+0.2$ for
$\Delta=0,\,r_{c}\approx 15$.
The plots in Figs. 3-5 indicate transition of the attractor from the
invariant torus (Fig. 3) to the Lorenz strange attractor (Fig. 5) as $\Delta$
decreases.

\item[Fig. 6] Fourier spectra $|X(\omega)|$ and
$|Z(\omega)|$ at $\sigma=5,\,b=1$ and $r=r_{c}+0.2$
for $\Delta=0.5$. There are three peaks in the spectrum $|X(\omega)|$:
$\omega=0$, $\omega\approx 0.04$ (see the inset in the upper right corner),
$\omega\approx \alpha=5.5$. There is no low frequency splitting mode in the
spectrum $| Z(\omega)|$, so that the second peak is associated with the
Goldstone-type mode (see Sec. 3).

\item[Fig. 7] Fourier spectra $|X(\omega)|$ and
$|Z(\omega)|$ at $\sigma=5,\,b=1$ and $r=r_{c}+0.2$
for $\Delta=0.1$. Both spectra are typical of doublings.

\item[Fig. 8] The irregular (noisy) Fourier spectrum $|X(\omega)|$
 at $\sigma=5,\,b=1$ and $r=r_{c}+0.2$ for $\Delta=0.05$.

\end{description}
\newpage

\epsfysize=200mm
\epsffile{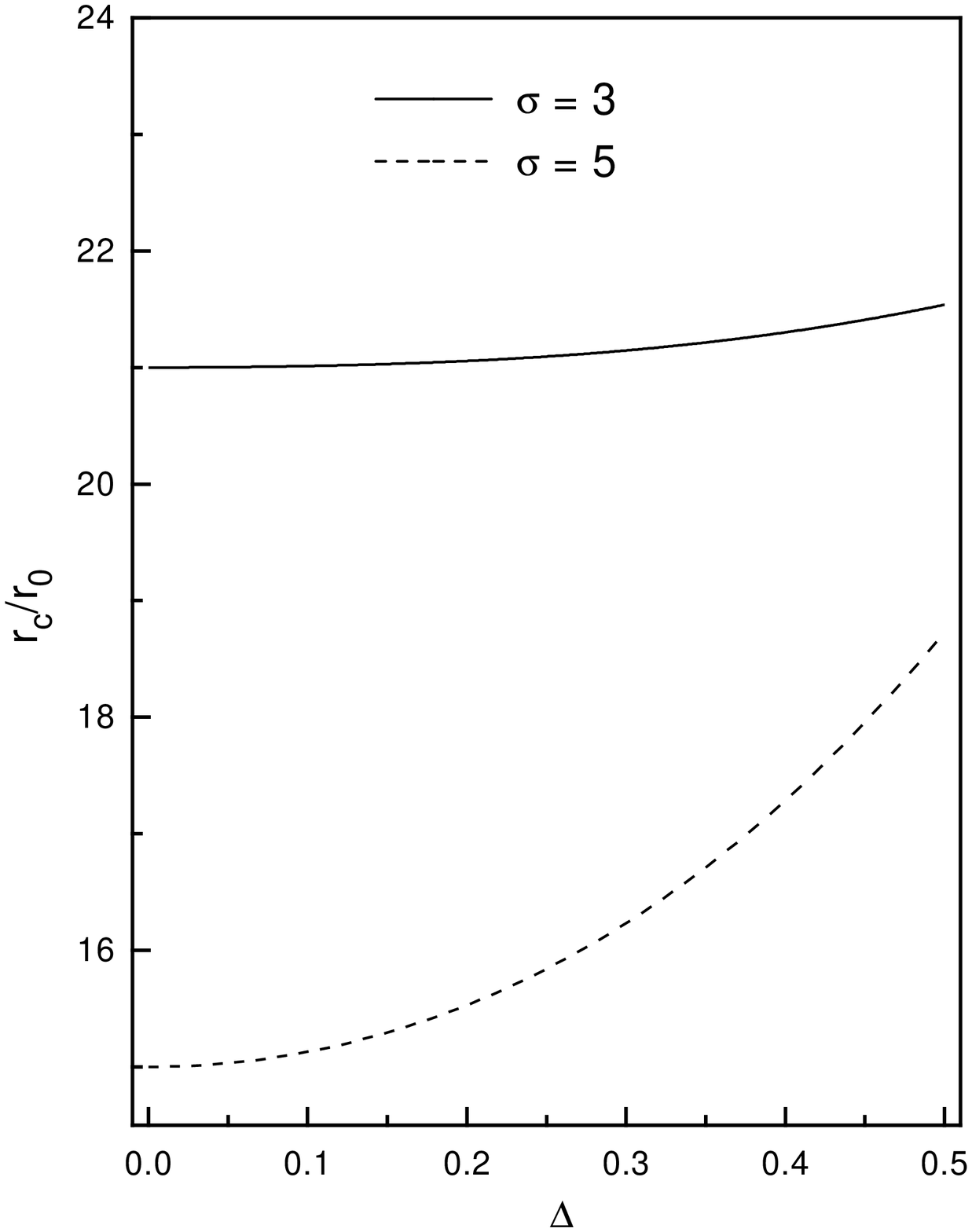}

\begin{center}
\bf FIGURE 1
\end{center}
\newpage

\epsfysize=200mm
\epsffile{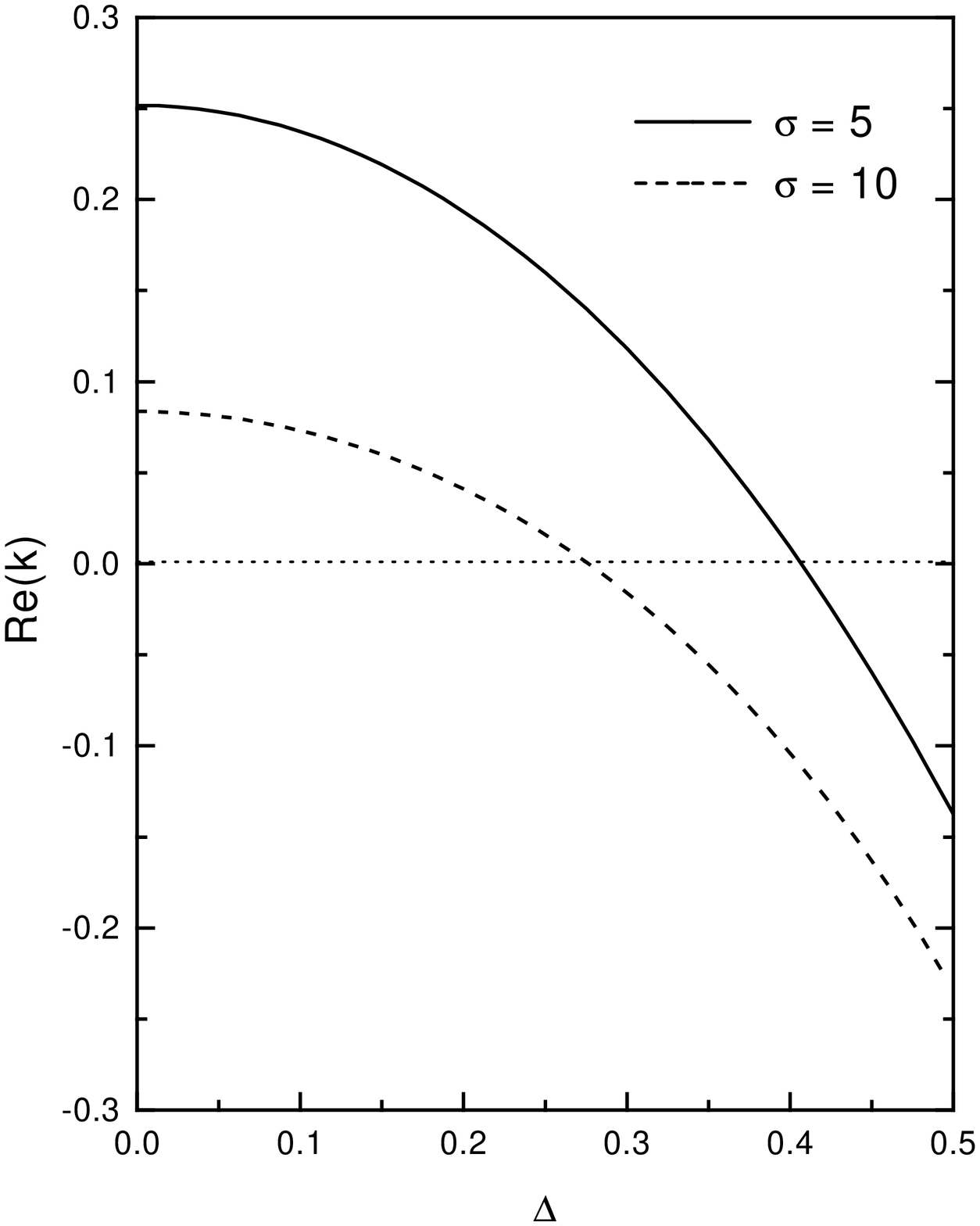}

\begin{center}
\bf FIGURE 2
\end{center}
\newpage

\epsfysize=200mm
\epsffile{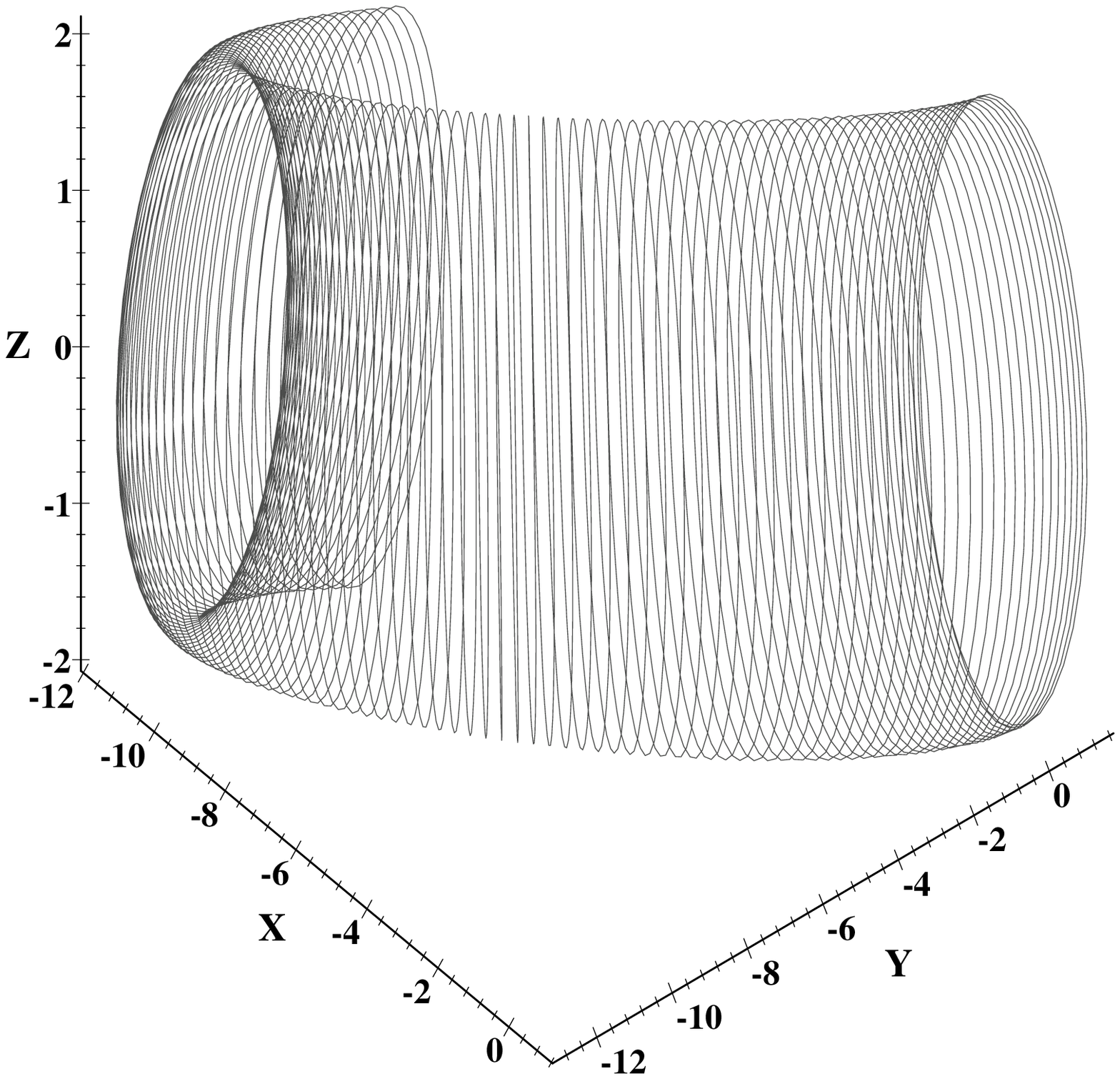}

\begin{center}
\bf FIGURE 3
\end{center}
\newpage

\epsfysize=200mm
\epsffile{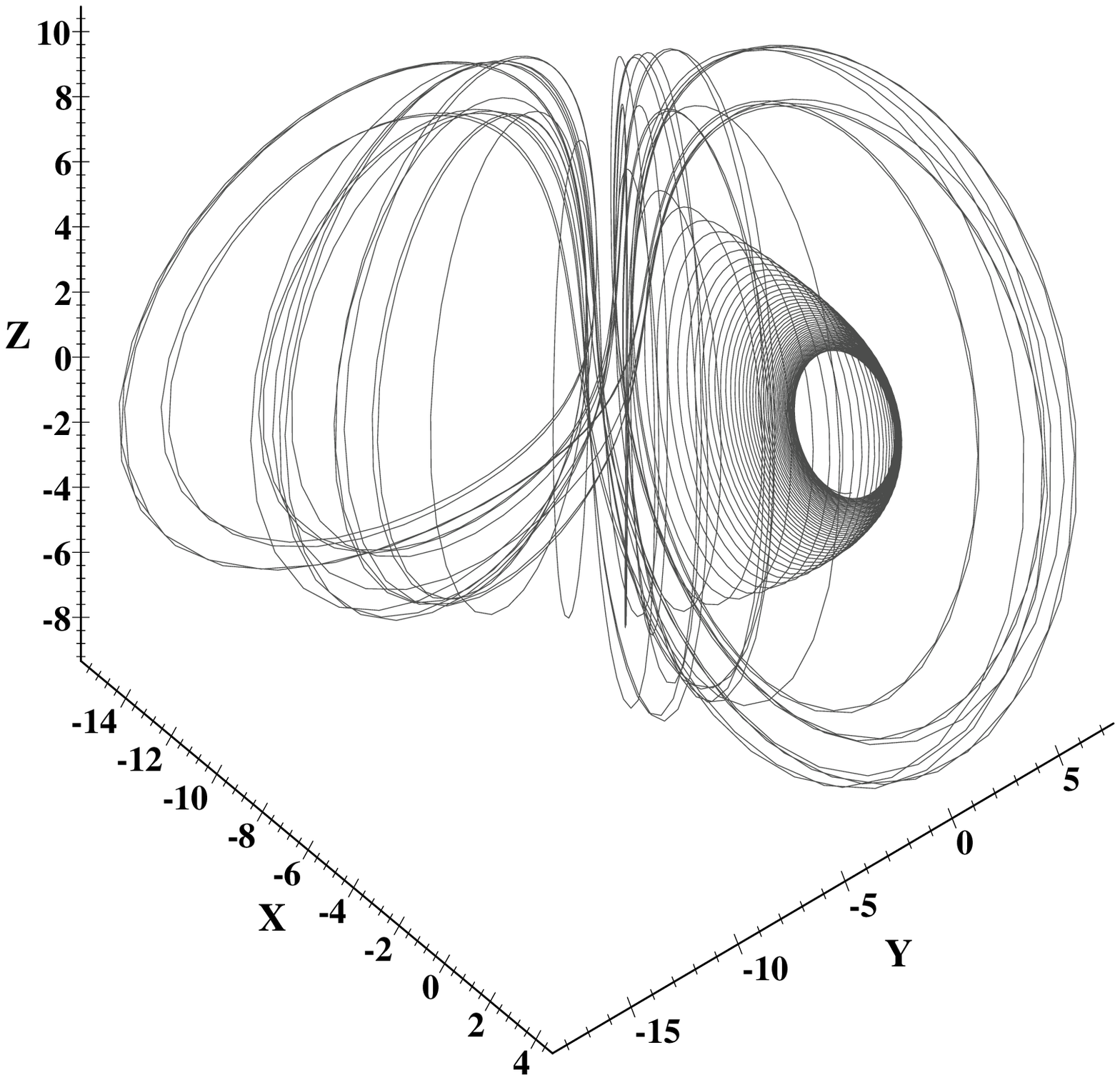}

\begin{center}
\bf FIGURE 4
\end{center}
\newpage

\epsfysize=200mm
\epsffile{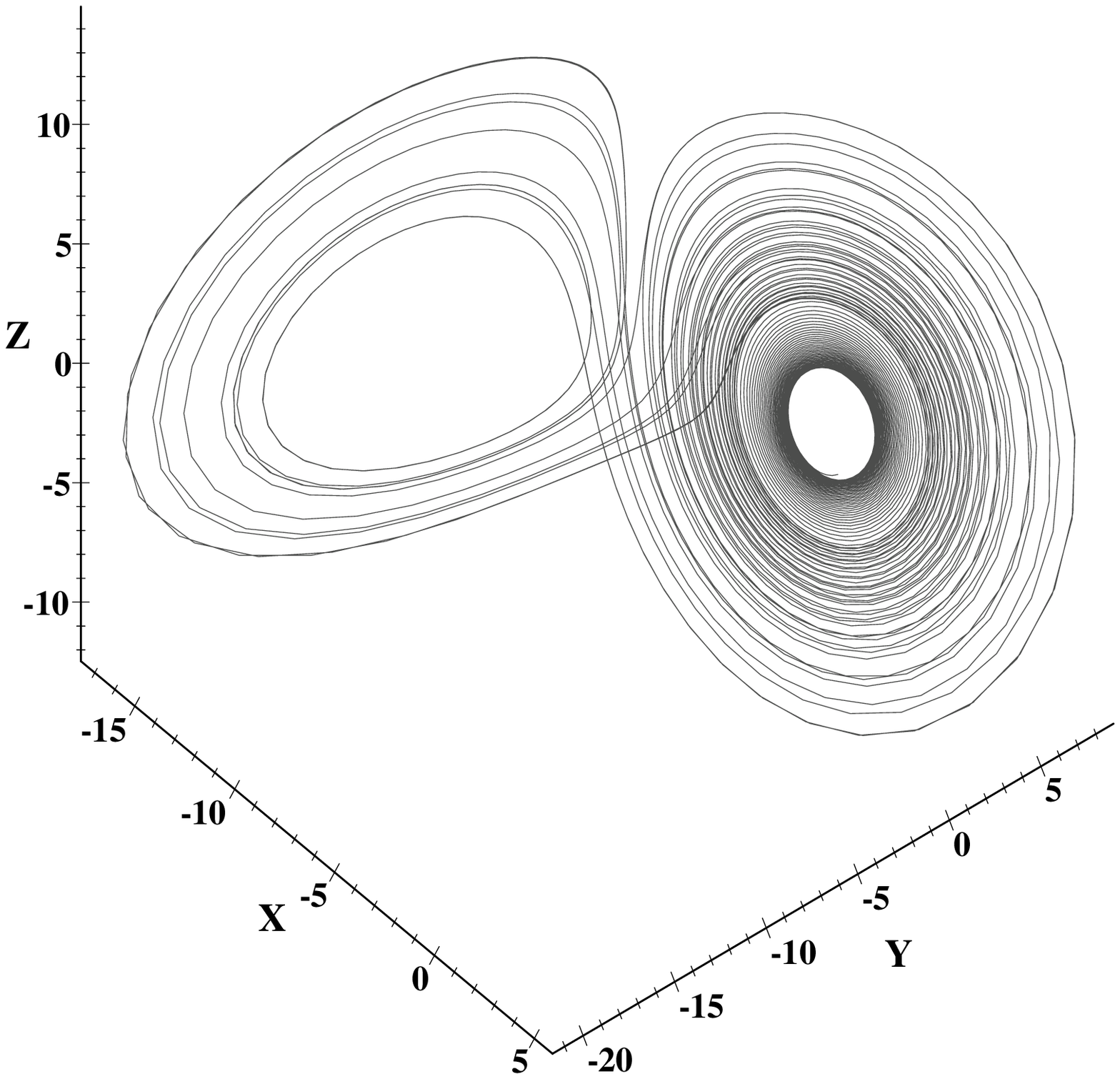}

\begin{center}
\bf FIGURE 5
\end{center}
\newpage

\epsfysize=200mm
\epsffile{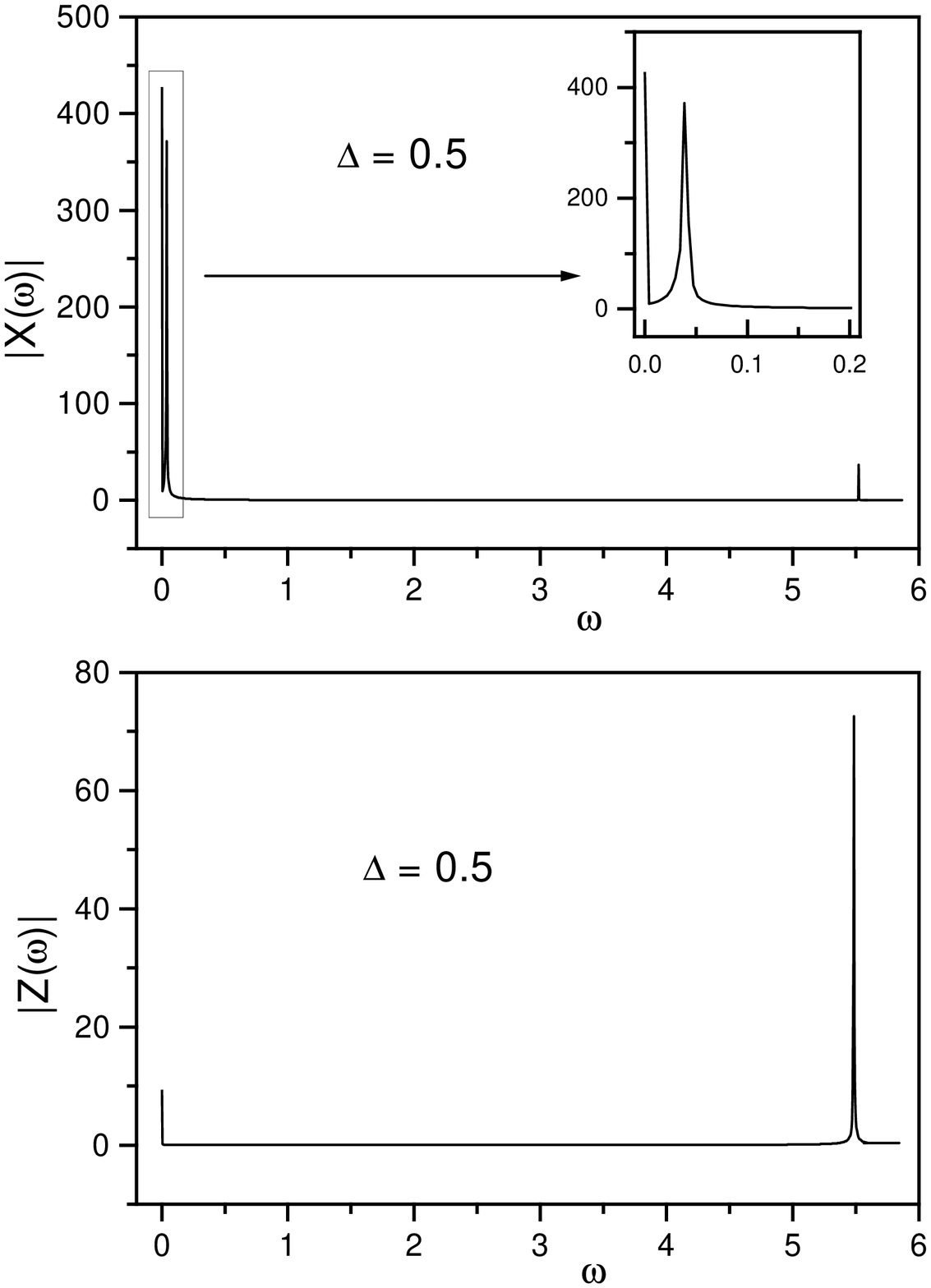}

\begin{center}
\bf FIGURE 6
\end{center}
\newpage

\epsfysize=200mm
\epsffile{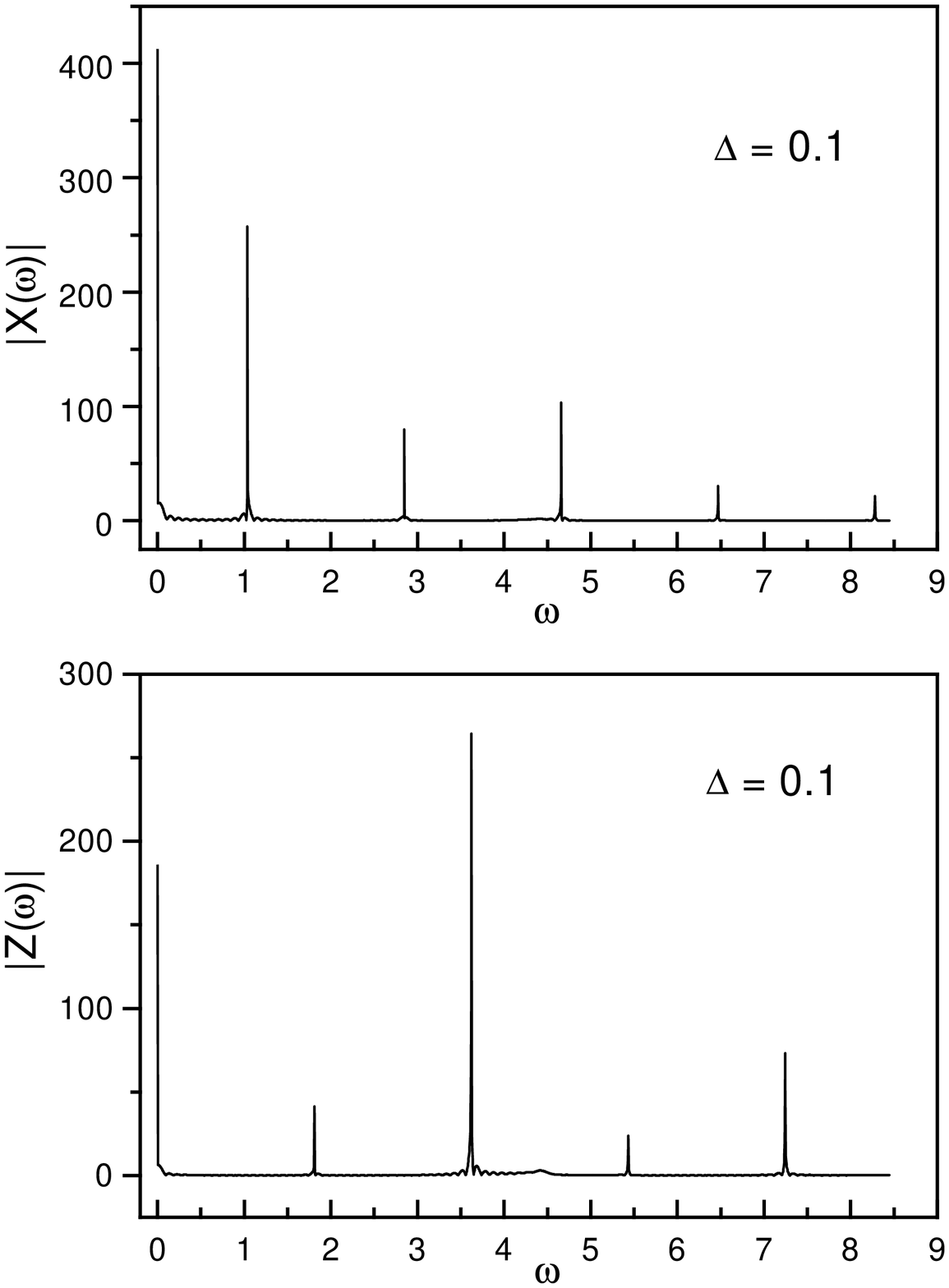}

\begin{center}
\bf FIGURE 7
\end{center}
\newpage

\epsfysize=200mm
\epsffile{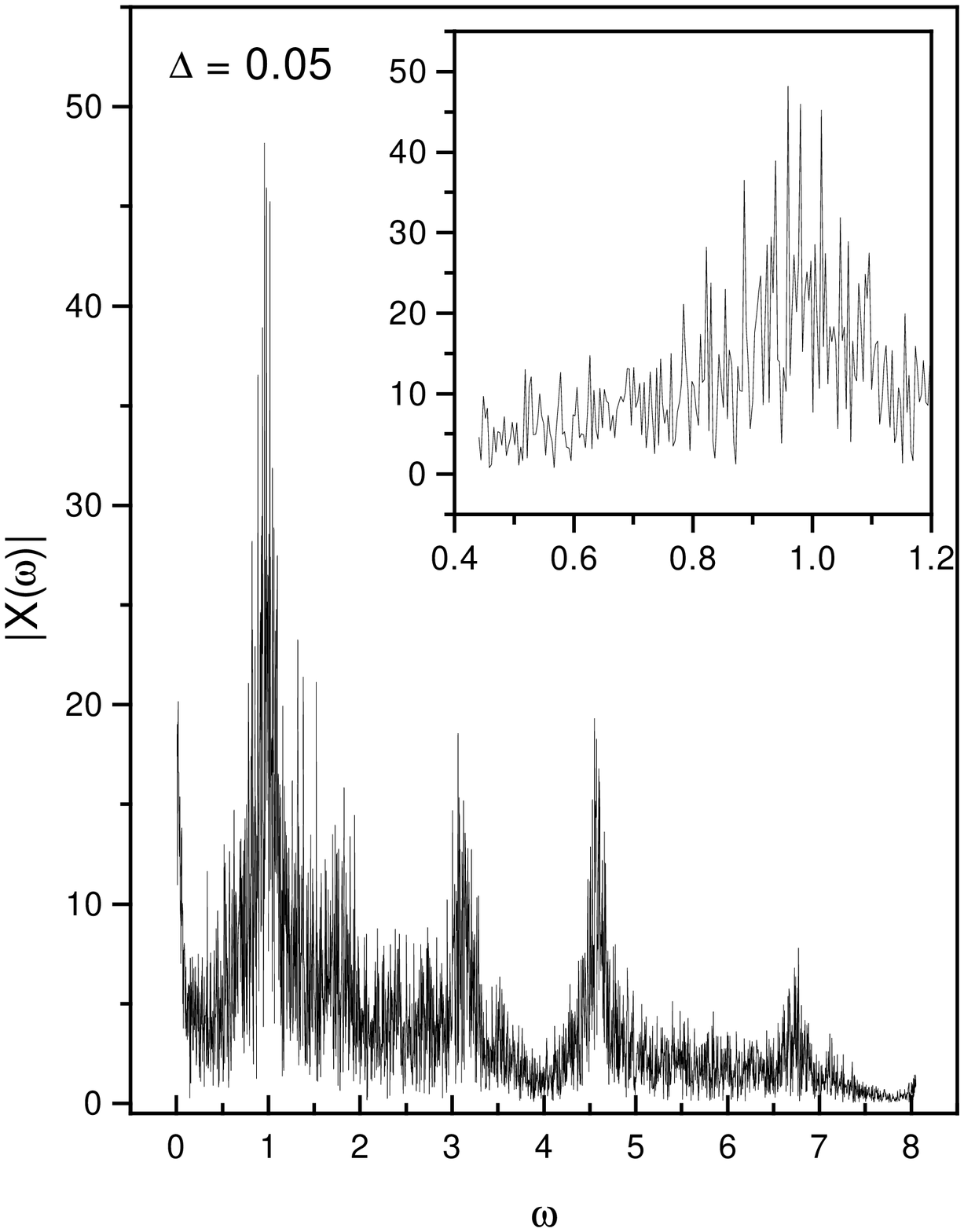}

\begin{center}
\bf FIGURE 8
\end{center}

\end{document}